\documentstyle[stwol]{article}

\input{psfig}

\def\Journal#1#2#3#4{{#1} {\bf #2}, #3 (#4)}

\def\PRD{{\em Phys. Rev.} D}

\def\roughly#1{\mathrel{\raise.3ex\hbox{$#1$\kern-.75em\lower1ex
\hbox{$\sim$}}}}

\def\be{\begin{equation}}
\def\ee{\end{equation}}
\def\bea{\begin{eqnarray}}
\def\eea{\end{eqnarray}}
\def\i3p{{I'_3}}
\def\darr{\raise1.5ex\hbox{$\leftrightarrow$}\mkern-16.5mu \partial_\mu}
\def\diag{{\rm diag}} 
\def\sign{{\rm sign}} 
\def\sss{\scriptscriptstyle}
\def\SM{{\scriptscriptstyle SM}}
\def\sfrac#1#2{{\scriptstyle #1/#2}}

\begin{document}

\title{$R_b$ and New Physics: A Comprehensive Analysis}

\author{\vspace{-1in}
\rightline{McGill/96-35}
\rightline{hep-ph/9609327}
\vspace{0.5in}}

\author{J.M.~CLINE}

\address{McGill University, 3600 University St., Montr\'eal, Qu\'ebec
H3A 2T8, Canada}

\twocolumn[\maketitle\abstracts{ I summarize work\cite{BBCNL} done with
Bamert, Burgess, Nardi and London in which we examine the effects of
new physics on $R_b$.  Our work facilitates the identification of such
new physics in case the present two-sigma deviation of $R_b$ from its
Standard Model value should persist, or alternatively it helps to
constrain new physics in case the discrepancy continues to disappear.
We consider two general classes of effects: mixing of the $b$ or $t$
quarks with heavy, exotic quarks, and loop corrections to the $Zb\bar
b$ vertex.  Our aim was to give an analysis which is as exhaustive and
as model-independent as possible.}]

\section{The (late?) $R_b$ problem}

As of one day before this talk was given, there appeared to be a
serious discrepancy with the LEP measurement of $R_b$, the ratio of the
partial widths $\Gamma(Z\to b\bar b)$ versus $\Gamma(Z\to {\rm
hadrons})$.  Naturally, this inspired a plethora of possible
theoretical explanations.  In trying to judge the relative merits of
one extension of the Standard Model versus another, it would be useful
to have some kind of unified framework in which to view them, so as to
be able to answer questions like ``how special is this particular
model?'' or ``could another similar model do the job better?'' \ We
were thus led to consider the more general question ``What kind of new
physics can increase the $Zb\bar b$ vertex?'' and to try to answer it
in as model-independent a way as possible.

Let me first review the experimental situation as it stood just after the
Moriond 1996 conference.\cite{lep}  Normalizing the $Zb\bar b$ couplings as
\begin{equation}
  \frac{e}{\sin2\theta_{\sss W}}\bar b \gamma^\mu\left( g^b_L (1-\gamma_5) + 
  g^b_R (1+\gamma_5) \right) b\, Z_\mu,
\label{bcouplings}
\end{equation}
their Standard Model values can be expressed as 
\begin{eqnarray}
g^b_L &=& -\frac12 + \frac13\sin^2\theta_{\sss W} = -0.4230;\nonumber\\
g^b_R &=&  + \frac13\sin^2\theta_{\sss W} = +0.770. 
\label{smcouplings}
\end{eqnarray}
Allowing for deviations $\delta g^b_L$ and  $\delta g^b_R$ of these
couplings from their Standard Model values, possibly due to new
physics, and performing a global fit to the $Z$-pole data, one obtains
the results\cite{peter} summarized in table 1.   Column 1 shows which
parameters among $\delta g^b_L$ and  $\delta g^b_R$ are allowed to
vary, so that row 1 (where neither vary) represents the
Standard Model, which is a poor fit to the data. This is due to the
measured $R_b = 0.2219\pm 0.0017$ having been larger than the Standard
Model value of $R_b^\SM=0.2156$. Allowing $\delta g^b_L$ or $\delta
g^b_R$ to vary increases the confidence level of the fit from 2.5\% to
$26-37$\%, a quite substantial improvement.  From a statistical point
of view, it is sufficient to adjust either of the couplings, not
necessarily both, to get an acceptable fit. 

Table 1 shows that it takes a much bigger relative change in the
right-handed than the left-handed coupling to change $R_b$ by the
amount that was needed:  $\delta g^b_R/g^b_R\sim 40$\%  versus $\delta
g^b_L/g^b_L\sim 2$\%.  This immediately tells us that we need
tree-level physics, for example mixing of $b_R$ with new quarks, to get
a big enough change in $g^b_R$.  On the other hand, one-loop effects
{\it or} small tree-level mixing can give a sufficiently large change
in the left-handed coupling.  There is a further possibility we do not
discuss, but which has been pursued by others,\cite{mangano} mixing of
the $Z$ boson with a $Z'$.

\begin{table}\begin{center}\caption{Global fits to Moriond $Z$-pole 
data.$^3$}
\vspace{0.4cm}
\begin{tabular}{|c|c|c|c|c|} 
\hline 
vary & $\delta g^b_L$ & $\delta g^b_R$ & c.l. of fit & $\chi^2/$d.o.f \\ 
\hline 
none & 0 & 0 & 2.5\% & $24.7/13$ \\

$\delta g^b_L$ & ${-0.0063\atop\pm 0.0020}$ & 0 & 26\% & $14.7/12$ \\

$\delta g^b_L$ & 0 & ${0.034\atop\pm 0.010}$ & 37\% & $13.0/12$ \\

both & ${-0.0029\atop\pm 0.0037}$ & ${0.022\atop \pm 0.018}$ & 32\% & $12.5/11$ \\
\hline
\end{tabular}
\end{center}
\end{table}

\section{Quark mixing}\subsection{$b$-$b'$ mixing} 

A simple way to change the $Zb\bar b$ couplings is to allow mixing
between the $b$ and a new exotic $b'$ with both left- and right-handed
components.  The couplings become $g^b_{\sss L,R} \to
\cos^2\theta_{\sss L,R}\, g^b_{\sss L,R} + \sin^2\theta_{\sss L,R}\,
g^{b'}_{\sss L,R}$, where the mixing angles come from the similarity
transformation $M\to O^T_L M O_R$ needed to diagonalize the $2\times 2$
Dirac mass matrix $M$ for $b$ and $b'$.  We were able to simplify the
range of possibilities by making three reasonable assumptions.  First
of all one must assume that $M_{12}$ or $M_{21}$ is nonvanishing in
order to have any mixing at all.  Second, we expect the $b'$ to be
heavy, which is only possible if $M_{22}$, the entry corresponding to
the $b'$ mass in the case of zero mixing, is large.  Finally we exclude
Higgs boson representations higher than doublets, since these tend to
make undesirably large contributions to the rho parameter.

Under these assumptions there are only twelve possible isospin
assignments for the chiral components of the $b'$ quark, which are
listed in Table 2, along with the mixing angle needed ($s^2_{\sss L,R}
\equiv \sin^2\theta_{\sss L,R}$) to change $g^b_L$ or $g^b_R$ by the desired
amount.  In almost all of these models, only one of the two mixing
angles is relevant, the other one being suppressed by powers of the
first.  However for the model with $I'_L = I'_{3L} = 0$ and $I'_R = 
- I'_{3R} = 1/2$, a combination of both left- and right-handed mixing
angles is possible, illustrated in figure 1.  

\begin{table}
\caption{Twelve models of $b$-$b'$ mixing that solve the $R_b$ problem,
labeled by the isospins of the $b'_L$ and $b'_R$.}
\vspace{0.4cm}
\begin{center}
\begin{tabular}{|c|c|c|c|c|}
\hline
$I'_L$ & $\i3p_L$ & $I'_R$ & $\i3p_R$ & {angle needed}\\
\hline
$1$ & $-1$&$1$ & $-1$ &  
$s_L^2=0.0111\pm 0.0032$\\
$1$ & $-1$&$\sfrac12$ & $-\sfrac12$& " \\
$\sfrac32$ & $-\sfrac32$&$1$ & $-1$& $s_L^2=0.0056\pm 0.0016$\\
$\sfrac12$ & $\sfrac12$&$\sfrac12$ & $\sfrac12$&
 $s_R^2=0.052 {+ 0.013\atop -0.014}$\\
$0$ & $0$&$\sfrac12$ & $\sfrac12$&"\\
$\sfrac12$ & $\sfrac12$&$1$ & $1$& $s_R^2=0.026 {+ 0.006\atop -0.007}$\\
$\sfrac12$ & $\sfrac12$&$0,1$ & $0$& $s_L^2=0.8515\pm 0.0016$\\
$\sfrac32$ & $\sfrac12$&$1$ & $0$&"\\
$0$ & $0$&$\sfrac12$ & $-\sfrac12$& $s_R^2{\roughly >} 0.361$\\
$\sfrac12$ & $-\sfrac12$&$\sfrac12$ & $-\sfrac12$&$s_R^2=0.361 
{+ 0.013\atop -0.014}$ \\
$\sfrac12$ & $-\sfrac12$&$1$ & $-1$&$s_R^2=0.180\pm 0.007$\\
\hline \end{tabular} \end{center}
\end{table}

\begin{figure}
\psfig{figure=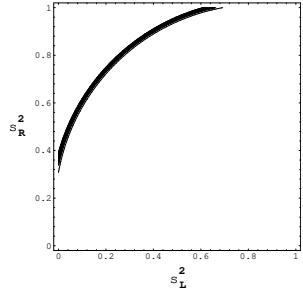,height=1.5in}
\caption{Allowed values of mixing angles in the $s^2_R$-$s^2_L$ plane for
model number 10 of table 2.}
\label{fig:fig1}
\end{figure}

\subsection{$t$-$t'$ mixing}

\begin{figure}
\psfig{figure=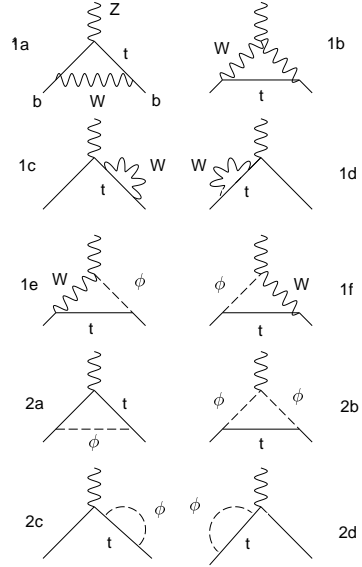,height=3in}
\caption{Top quark loop diagrams that affect $g^b_L$. }
\label{fig:tp}
\end{figure}

Another possibility is that the top quark mixes with an exotic $t'$,
which alters the $g^b_L$ coupling through the loop diagrams of figure
\ref{fig:tp}.  These are the same as the diagrams of the Standard
Model, except that the top quark must be replaced by two linear
combinations of $t$ and $t'$, one for each chirality.  Making the same
assumptions about the $t$-$t'$ mass matrix as we did for that of $b$
and $b'$ above, there are again twelve possible isospin assignments for
the $t'$ quark, enumerated in Table 3.

\begin{table}
\caption{Twelve possible models of $t$-$t'$ mixing,
labeled by the isospins of the $t'_L$ and $t'_R$.}
\vspace{0.4cm}
\begin{center}
\begin{tabular}{|c|c|c|c|}
\hline
$I'_L$ & $\i3p_L$ & $I'_R$ & $\i3p_R$ \\
\hline
 $3 / 2$ & $+{1 / 2}$ & 1 & $ +1 \> $\\
 $1 / 2$ & $+{1 / 2}$ & 1 & $ +1,0 \> $\\
 $1 / 2$ & $+{1 / 2}$ & $1 / 2$ & $+{1 / 2} $\\
 $1 / 2$ & $+{1 / 2}$ & $ 0 $ & $ 0 $ \\
 $1 / 2$ & $-{1 / 2}$ & 1 & $ 0,-1 $ \\
 $1 / 2$ & $-{1 / 2}$ & $1 / 2$ & $-{1 / 2} $\\
 $1 / 2$ & $-{1 / 2}$ & $ 0 $ & $ 0 $\\
 0 & 0 & $1 / 2$ & $\pm{1 / 2} $\\
 0 & 0 & $ 0 $ & $ 0 $  \\
\hline \end{tabular} \end{center}
\end{table}

The resulting change in $g^b_L$ can be expressed rather simply in the
limit where both top quarks are much heavier than the $W$ boson, if we
concentrate on values of $m_{t'}$ so as to maximize the magnitude of 
$g^b_L$.  This turns out to occur when $m_{t'}\ll m_t$, in which case a
single term dominates the shift in $g^b_L$ beyond its Standard Model
value,
\begin{equation} 
\delta g^b_L \cong {\alpha\over 16\pi s^2_w} V^2_{t'b}\left(
{m^2_{t'} - m^2_t\over m^2_W} + 6\ln{m_{t'}\over m_t}\right).
\end{equation}
Here $V_{t'b}$ is the element from the extended CKM matrix which
includes the $t'$ quark in addition to $u$, $c$ and $t$.  The best case
is when $m_{t'}$ is at its experimental lower limit of 135 GeV, in
which case $\delta g^b_L = -0.0021$.  Although this is too small to
explain the previous LEP data which needed $\delta g^b_L = -0.0063$,
the recent results of ALEPH announced at this conference\cite{Tomalin}
have reduced the size of the discrepancy between theory and the world
average measurement\cite{monig} of $R_b=0.2178\pm 0.0011$ such that $-0.0021$ 
is now sufficient.

\section{General Loop Effects}\subsection{Diagonal $Z$ Couplings}

\begin{figure}
\psfig{figure=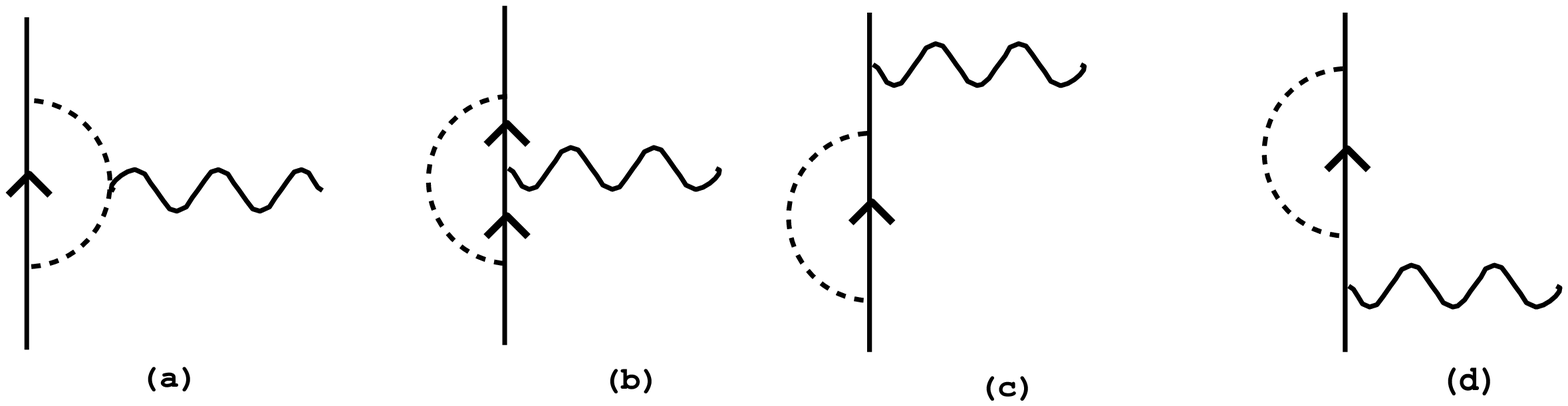,width=3in}
\caption{The important generic loops for $g^b_L$}
\label{fig:loops}
\end{figure}

\begin{figure}
\psfig{figure=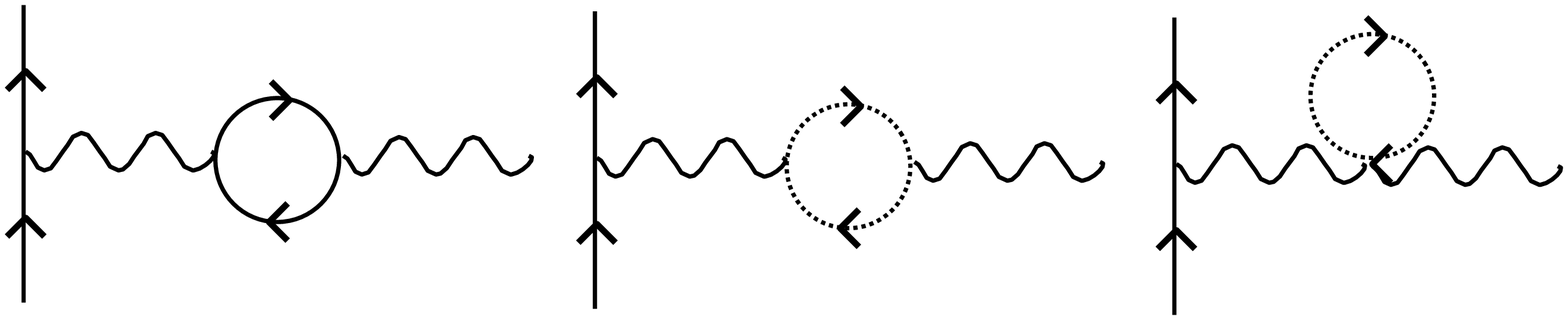,width=3in}
\caption{Vacuum polarization contribution which tend to cancel in $R_b$.}
\label{fig:vacpol}
\end{figure}

In the previous subsection we had to consider loops involving the $t'$
as a special case because they were inextricably tangled up with usual
Standard Model contributions, but here we want to consider the effects
of a general fermion $f$ and scalar $\phi$ coupling to the $b_L$ via the
interaction ${\cal L}_y = y \bar b_L\phi f +$ h.c., and to the $Z$ boson
via  
\begin{equation}
{2e\over\sin2\theta_W}\left(\bar f\gamma_\mu(g^f_L P_L+
g^f_R P_R)f + ig^\phi\phi^\dagger\darr\phi\right) Z^\mu.
\label{Zcouplings}
\end{equation}
The relevant diagrams which modify $g^b_L$ are shown in figure
\ref{fig:loops}.  There are also vacuum polarization diagrams as in
figure \ref{fig:vacpol}, but these cancel to a good approximation in the
ratio that defines $R_b$.  The result of evaluating the first four
diagrams has a remarkably simple expression in the approximation of
ignoring $M_Z$:
\begin{equation}
\delta g^b_L = {y^2 n_c\over 16\pi^2}(g^f_L-g^f_R){\cal
F}(m^2_f/m^2_\phi),
\label{diag}
\end{equation}
where $n_c$ is a color factor that is unity if $\phi$ is colorless and
$f$ is a color triplet, for example.  The kinematic function given by
${\cal F}(x) = x/(x-1) - x\ln x/(x-1)^2$ is always positive and reaches
its maximum value of 1 when $m_f/m_\phi\to\infty$.  The $M_Z=0$
approximation turns out to be better than one would expect, since the
first order correction in powers of $M^2_Z/m^2_f$ is typically only
10\% of eq.~(\ref{diag}), even when $M^2_Z/m^2_f=1$.

A notable feature of eq.~(\ref{diag}) is that its sign is completely
determined by the isospins of $f_L$ and $f_R$ since $g^f_L-g^f_R =
I^f_{3L} - I^f_{3R}$.  Thus one can immediately see why two-Higgs
doublet models, as well as the top quark contribution in the Standard
Model, tend to decrease $R_b$: if $f$ is the top quark then $I^f_{3L} -
I^f_{3R} = 1/2$, which has the opposite sign to the tree level value of
$g^b_L$, reducing the effective coupling.   In two-Higgs models with
very large $\tan\beta$, so that the bottom quark itself can make an
appreciable contribution in the loop, the sign is such as to increase
$R_b$.

\subsection{Nondiagonal $Z$ Couplings}

It might happen that the flavor states with the largest coupling to
$b_L$, which we shall now denote by $\phi_1$ and $f_1$, mix with other
flavor states $\phi_2$ and $f_2$ to form mass eigenstates $\phi,\phi'$
and $f,f'$. Then our previous formula (\ref{diag}) for $\delta g^b_L$
must be generalized to something more complicated.  Nevertheless in
certain limiting cases it becomes relatively simple again.  First, if
the masses are degenerate such that $m_f = m_{f'}$ and $m_\phi=m_{\phi'}$,
we get
\begin{equation}
\delta g^b_L = {y^2 n_c\over 16\pi^2}(U_R U^\dagger_L g^f_L U_L
U^\dagger_R -g^f_R)_{11}{\cal
F}(m^2_f/m^2_\phi).
\label{nondiag1}
\end{equation}
Here $g^f_{L,R}$ are the $2\times 2$ matrix generalizations of the
fermion couplings to the $Z$ boson, and $U^\dagger_L M U_R$ is the
similarity transformation that diagonalizes the $f_1$-$f_2$ mass
matrix.  Second, if the fermions are much heavier than the scalars,
we get eq.~(\ref{nondiag1}) again, with the addition of two extra terms
that tend to cancel each other, at least in the supersymmetric case we
will discuss below.  Third, if the scalars are much heavier than the
fermions we obtain 
\begin{equation}
\delta g^b_L = {y^2 n_c\over 16\pi^2}\sin^2 2\theta_\phi 
(g^\phi_{11}-g^\phi_{22}){\cal
F}'(m^2_f/m^2_\phi),
\label{nondiag2}
\end{equation}
where $\theta_\phi$ is the scalar mixing angle and ${\cal F}'(x)
={x+1\over 2(x-1)}
\ln x -1$ is
another positive function like ${\cal F}$.  Thus it is once again
rather easy to determine the sign and magnitude of $\delta R_b$.

An example is the supersymmetric Standard Model where, because of the
large top quark yukawa coupling, $\phi_1$ is the right-handed top
squark $\tilde t_R$, and $f_1$ is the higgsino $\tilde h_2^-$, or
alternatively the fermion and scalar are the usual top quark and one of
the Higgs bosons, $\phi_1 = h_2^-$ and $f_1 = t_R$.  We already
explained that the quark-Higgs contribution has the wrong sign for
increasing $R_b$, so one wants to minimize these loops by taking the
Higgs bosons to be very heavy.  Concentrating on the squark and
higgsino, therefore, one must take into account their mixing with the
left-handed top squark, $\phi_2=\tilde t_L$ and the Wino, $f_2 = \tilde
W^-$.  The charge matrices are $g^f_L = g^f_R = \diag(-1/2,-1)$ for the
fermions and $g^\phi=\diag(0,1/2)$ for the scalars.  Applying the
simplifying cases 1 or 2 mentioned above, one finds 
\begin{equation}
\delta g^b_L \cong - {y^2 n_c\over 32\pi^2}\sin^2(\theta_L
-\sign(m_f/m_{f'})\theta_R){\cal F},
\end{equation}
and therefore to have a large shift in $g^b_L$, the combination of
chargino mixing angles $|\theta_L -\sign(m_f/m_{f'})\theta_R|$ must be
large.  From this and the form of the chargino mass matrix it is easy
to deduce that both charginos must be rather light, and that
$\tan\beta$ must be close to unity.  Furthermore the third simplifying
limit above, that of heavy scalars, gives the wrong sign for increasing
$R_b$ because $g^\phi_{11} - g^\phi_{22} = -1/2$.  Thus one needs at
least one light scalar, which naturally enough turns out to be 
$\tilde t_R$ because this is the one that couples directly to the
left-handed $b$ quark.

\section{Summary} 

We have shown that a significant increase of $R_b$ from its Standard
Model value can be caused by new physics that changes $g^b_R$ by quark
mixing, or $g^b_L$ either by mixing or one loop contributions to the
$Zb\bar b$ coupling.  In the category of $b$-$b'$ mixing we found 12
possible models using modest restrictions.  The Standard Model $t$
quark loop corrections to $g^b_L$ can also be altered by $t$-$t'$
mixing, but only by as much as $-0.0021$, corresponding to a maximum
increase of $0.0017$ in $R_b$.

For more general kinds of loop corrections to $R_b$, we have found
simple approximate formulas which gives insight into the sign and
magnitude of the change.  Although the original motivation was to make
it easier to build models explaining the former discrepancy, perhaps
they will in the future be more useful in constraining models that
arise in other contexts.  It should however be kept in mind that the
world average value of $R_b$ as of this writing is still two standard
deviations higher than the Standard Model value, despite the recent
results of ALEPH which now agree with the Standard Model.


\section*{References}
\begin{thebibliography}{99}
\bibitem{BBCNL} P.~Bamert, C.P.~Burgess, J.M.~Cline, E.~Nardi and D.~London,
\Journal{\PRD}{54}{4275}{1996}.
\bibitem{lep}LEP electroweak working group,  
preprint CERN-PPE/95-172, 1995;
M.D. Hildreth, ``The Current Status of the $R_b$ and $R_c$ Puzzle,''
talk given at the XXXIst Rencontres de Moriond, March 1996.
\bibitem{peter} P.~Bamert, preprint  McGILL-95-64, hep-ph/9512445 (1996).
\bibitem{mangano} see for example M.~Mangano, ``Hadrophilic/ Leptophobic
$Z'$ Models and $R_b$,' in these proceedings.
\bibitem{Tomalin} I.~Tomalin, ``$R_b$ with Multivariate Analysis," in these
proceedings.
\bibitem{monig} K.~M\"onig, ``Combination 
of Heavy Quark E.W.~measurements,'' in these proceedings.
\end{thebibliography}
\end{document}